%% file: EntanglementStablizationMain.tex
\documentclass[aps,prl,amsmath,amssymb,twocolumn,groupedaddress,showpacs,superscriptaddress]{revtex4-2}

\input{preamble}

\begin{document}

\title{Hardware-Efficient Stabilization of Entanglement via Engineered Dissipation in Superconducting Circuits}

\input{author}
\date{\today}

\begin{abstract}

Generation and preservation of quantum entanglement are among the primary tasks in quantum information processing. State stabilization via quantum bath engineering offers a resource-efficient approach to achieve this objective. However, current methods for engineering dissipative channels to stabilize target entangled states often require specialized hardware designs, complicating experimental realization and hindering their compatibility with scalable quantum computation architectures. In this work, we propose and experimentally demonstrate a stabilization protocol readily implementable in the mainstream integrated superconducting quantum circuits. The approach utilizes a Raman process involving a resonant (or nearly resonant) superconducting qubit array and their dedicated readout resonators to effectively emerge nonlocal dissipative channels. Leveraging individual controllability of the qubits and resonators, the protocol stabilizes two-qubit Bell states with a fidelity of $90.7\%$, marking the highest reported value in solid-state platforms to date. Furthermore, by extending this strategy to include three qubits, an entangled $W$ state is achieved with a fidelity of $86.2\%$, which has not been experimentally investigated before. Notably, the protocol is of practical interest since it only utilizes existing hardware common to standard operations in the underlying superconducting circuits, thereby facilitating the exploration of many-body quantum entanglement with dissipative resources.
\end{abstract}

\maketitle

Quantum state preparation is typically achieved through coherent unitary operations\cite{nielsen_chuang_2010}. However, realistic quantum systems interact with the environment, resulting in decoherence that degrades their quantumness\cite{Arute2019,Wu2021prl,Zhong21PRL}. Although decoherence is typically seen as an obstacle to quantum information processing (QIP), it can, if controllable, offer distinct benefits over unitary-only circuits. For example, quantum bath engineering\cite{Harrington2022} has been shown to enable efficient processing with applications in stabilized state preparation\cite{Lin2013,Leghtas2015,Grimm2020,xiaomi2024}, quantum error correction (QEC)\cite{Gertler2023,Gertler2021,Xu2023}, and dissipation-assisted phase transition\cite{Naghil2019,Noel2022,Koh2023,han2023exceptional}. Consequently, bath engineering has evolved into a fruitful research field, encompassing applications such as cooling various quantum systems\cite{Schreck2021,Teufel2011,Kotler2021,Feng2020} and entanglement stabilization in both trapped ions\cite{Lin2013,Cole2022,Malinowski2022} and superconducting circuits\cite{Shankar2013,Schwartz2016,Brown2022}. This non-unitary technique for state stabilization is 
appealing because it directs a driven-dissipative system toward steady states of interest by engineering dissipative channels, making it insensitive to initial states and noise\cite{Malinowski2022}.

Circuit quantum electrodynamics (cQED), based on superconducting devices, constitutes a promising platform for QIP\cite{Blais2021}. In this platform, dissipative resonators coupled with superconducting qubits are employed to realize dispersive readout\cite{Kjaergaard2020}. Furthermore, when considered as generalized measurement, dissipation serves as a powerful quantum control tool that has been utilized in other QIP tasks within cQED\cite{Harrington2022,Kapit_2017}, including state stabilization and autonomous QEC in both single-qubit\cite{Murch2012,Geerlings2013,Yao2017,Magnard2018,Zhou2021,McEwen2021,Marques2023,lacroix2023fast} and two-qubit systems\cite{Shankar2013,Schwartz2016,Liu2016,Brown2022,Li2024}. This continuous driven-dissipative concept has been further generalized to many-body systems\cite{Ma2019,Petiziol2022prl,Yunzhao2023}, and specific eigenstates of a three-qubit array were prepared by Gourgy \textit{et al.}\cite{Gourgy2015}, although the quantum coherence of these states has yet to be confirmed. Nevertheless, in all existing experiments involving quantum bath engineering for entangled state stabilization, superconducting qubits were coupled to a common dissipative resonator requiring specialized hardware designs, rendering most of them incompatible with mainstream architectures for scalable quantum computation and simulation. Additionally, the stabilized Bell state fidelity achieved in cQED remains relatively low, with the highest reported value up to $84\%$\cite{Brown2022} so far, thus limiting its potential application in QIP. Therefore, a bath engineering scheme that is compatible with scalable architectures and capable of achieving high-fidelity entangled states in superconducting circuits is highly desirable.

In this Letter, we propose and experimentally implement a hardware-efficient scheme for stabilizing entangled states that is compatible with the mainstream and scalable architecture of contemporary superconducting quantum computation. We consider a driven-dissipative system of the transmon qubits\cite{Kochtramson} with their readout resonators. A Raman process \cite{Murch2012} is utilized by applying detuning drives to the resonators to tailor the spectrum of their quantum noise\cite{RevModPhys2010}, thereby engineering appropriate dissipative channels for generating steady and desired entangled states. Leveraging the individual addressing of qubits and resonators, we are able to stabilize a Bell state of two qubits with a fidelity of $90.7\%$ as determined by quantum state tomography (QST). The protocol is further extended to a three-qubit scenario, stabilizing an entangled $W$ state with a fidelity of $86.2\%$. While theoretically, these entangled states can be maintained indefinitely, the stabilization period demonstrated in our experiment is finite due to the limited buffer time of the apparatus. Our results demonstrate the potential of harnessing dissipation with hardware efficiency to stabilize multi-qubit entangled states.

\begin{figure}[htbp]
    \centering
    \includegraphics{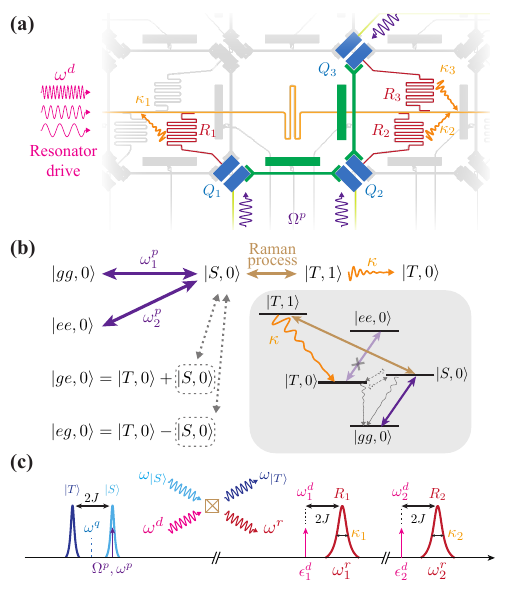}
    \caption{(a) Photograph of the superconducting circuits employed in the experiment, with the involved components highlighted. Each qubit $Q_i$ is paired with a dedicated readout resonator $R_i$. The coupling between qubits is adjusted using tunable couplers (green). (b) Main panel: illustration of the evolution of various initial states to the target Bell state of two qubits. The essential component of the flow is a Raman process. Numbers in the Dirac notation indicate Fock states of the resonators. Shaded panel: Energy level diagram of the system showing the Raman process (brown line), pumping drives applied to the qubits (purple line), and decoherence processes (yellow and gray wavy lines). The pumping drives selectively excite the transition from the ground state to $\ket{S}$ while inhibiting the transition $\ket{T}\leftrightarrow\ket{ee}$ due to symmetry of the pumping drives. The combination of the Raman process and rapid energy dissipation in the resonators stabilizes the system into the target Bell state $\ket{T}$. Only the first two energy levels of the resonator are shown for clarity. (c) Frequency landscape of the two resonators and the two Bell states, along with microwave pulses applied to all components. The Raman process is activated by injecting red-detuned ($\omega^r_i-\omega^d_i=2J$) drives into the resonators. The pumping drive is centered at $\omega_{\ket{S}}$, targeting the state $\ket{T}$. The inset illustrates the Raman process as four-wave mixing. } 
    \label{fig1}
\end{figure}

Our protocol is experimentally implemented on a superconducting quantum processor consisting of a 2D array of $4\times4$ frequency-tunable transmon qubits, with neighboring qubits coupled via frequency-tunable couplers\cite{feiyancpl}. The components involved in the protocol are highlighted in false color in \cref{fig1}(a). Each of the three transmon qubits $Q_i$, with a frequency of $\omega_i^{q}$, is dispersively coupled to a dedicated readout resonator $R_i$ with a frequency $\omega_i^r$, via a cross-Kerr coupling of strength $\chi_i$. The exchange coupling strength $J_{i,i+1}$ between $Q_i$ and $Q_{i+1}$ can be modulated by the relevant coupler. For all qubits, a common low-quality Purcell filter is designed to protect them from external environmental noise while providing moderate dissipation rates $\kappa_i$ for their readout resonators. The effective interaction Hamiltonian of the system, with near-resonant drives applied to both qubits and resonators, can be written as:
\begin{equation}
    \begin{aligned}
        H/\hbar=&\sum_{i=1}^L \Delta_i^q b_i^\dagger b_i+ \frac{\alpha_i}{2} b_i^\dagger b_i^\dagger b_i b_i+\Delta_i^r c_i^\dagger c_i +\chi_i b_i^\dagger b_i c_i^\dagger c_i \\
        -&\sum_{i=1}^{L-1}J_{i,i+1}(b^\dagger_ib_{i+1}+b_{i+1}^\dagger b_{i})\\
        +&\sum_{i=1}^L \Omega_i^p b_i^\dagger+\epsilon^d_i c_i^\dagger +h.c.,
    \end{aligned}\label{eq1}
\end{equation}
Here, $b_i^\dagger$ and $c_i^{\dagger}$ are the creation operators for $Q_i$ and $R_i$, respectively, and $\alpha_i$ denotes the anharmonicity of $Q_i$. All qubits are pumped by microwave pulses with the same frequency of $\omega^p$ ($\Delta_i^q=\omega_i^q-\omega^p$ denotes the corresponding detuning) but different amplitudes of $\Omega_i^p$. Both $\omega^p$ and $\Omega_i^p$ can be tuned \textit{in situ}. The Raman process is realized by driving $R_i$ using a microwave pulse with a frequency of $\omega_{i}^d$ ($\Delta_i^r=\omega_i^r-\omega_{i}^d$) and an amplitude of $\epsilon_i^d$ (see Supplemental Materials\cite{supplement} (SM) for experimental parameters).

We first demonstrate the generation of the Bell states of $Q_1$ and $Q_2$ using our dissipative scheme, corresponding to $L=2$ in \cref{eq1}. Consider the situation where all microwave drives on qubits and resonators are turned off, and the two qubits are tuned into resonance at $\omega^q=\omega_1^q=\omega_2^q$ with a moderate coupling strength of $J/2\pi=J_{12}/2\pi=5$ MHz. It can then be easily verified that the two Bell states, $\ket{T}=(\ket{ge}+\ket{eg})/\sqrt{2}$ and $\ket{S}=(\ket{ge}-\ket{eg})/\sqrt{2}$, become the eigenstates of the single-excitation manifold, with the difference between their eigenenergies satisfying $\omega_{\ket{S}}-\omega_{\ket{T}}=2J$. Here, $g(e)$ denotes the ground (excited) state of the qubits. The energy level diagram of the system is illustrated in \cref{fig1}(b). Although these Bell states can be coherently prepared through a combination of single- and two-qubit gates, such unitary processes are prone to noise and control errors. Moreover, the entangled states prepared in this manner are constantly degraded by both $T_2$ (dephasing) and $T_1$ (relaxation) processes after preparation. Furthermore, since the conventional method of dealing with noise through measurement-based QEC is resource-demanding\cite{LiuPRX,Andersen2019,Bultinksd,Ristè2020}, quantum bath engineering has been developed to circumvent this issue due to its inherent autonomous feedback loop\cite{Harrington2022,Kapit_2017}. 

Figure \ref{fig1}(b) shows an example of our scheme where the $\ket{T}$ state is the target Bell state. The key concept involves properly tuning the energy landscape of the system and applying appropriate microwave drives to design the transitions and dissipation channels to work collaboratively, steering any initial state towards the target Bell state. Moreover, this steering flow dynamically and autonomously stabilizes the Bell state against decoherence. The critical component within this steering flow is a Raman process that drives the $\ket{S,0}$ state to the $\ket{T,1}$ state. This process leverages the rapid dissipation in the readout resonators, as exploited in pioneering works on cooling single and multiple qubits in 3D superconducting cavities \cite{Murch2012,Gourgy2015}. This Raman scattering can be intuitively understood as a four-wave mixing process involving the transmons and their resonators $R_i$, as depicted in the inset of \cref{fig1}(c). With microwave drive(s) applied to one or both resonators at a frequency $\omega^d_i$ satisfying $\omega^r_i-\omega^d_i = \omega_{\ket{S}}-\omega_{\ket{T}}$, the input $\ket{S,0}$ state is scattered into the $\ket{T,1}$ state, with the gained energy populating the resonator(s) and dissipating efficiently. Consequently, the $\ket{T,1}$ state rapidly decays into the target $\ket{T,0}$ state. Evidently, such scattering occurs most efficiently when $\omega^r_i-\omega^d_i = \omega_{\ket{S}}-\omega_{\ket{T}} = 2J$, where the anti-Stokes photons resonate with $R_i$ due to the cross-Kerr coupling $\chi_i b_i^\dagger b_i c_i^\dagger c_i$.

To maintain the system in the target state $\ket{T,0}$, it is necessary to counterbalance the continuous relaxation from the subspace of $\ket{S,0}$ and $\ket{T,0}$ to the ground state $\ket{gg,0}$. This can be achieved by concurrently driving both qubits with pulses of the same frequency $\omega^p$ but opposite amplitudes, i.e., $\Omega_1^p = -\Omega_2^p = \Omega^p$, where $\Omega^p/2\pi = \SI{0.53}{MHz}$. Here, $\omega^p$ is chosen to ensure a resonant transition between $\ket{gg}$ and $\ket{S}$ while strongly suppressing the $\ket{S}\leftrightarrow\ket{ee}$ transition as $\Omega^p \ll 2J$. Meanwhile, the $\ket{T}\leftrightarrow\ket{ee}$ transition is forbidden due to a dynamical symmetry\cite{supplement}. When these continuous and concurrent drives are applied to the qubits, in combination with the previously discussed Raman process, the state $\ket{T,0}$ approximates a steady state of the system, even in the presence of decoherence. A similar scheme can be implemented for the other Bell state $\ket{S}$.

\begin{figure}[htbp]
    \centering
    \includegraphics{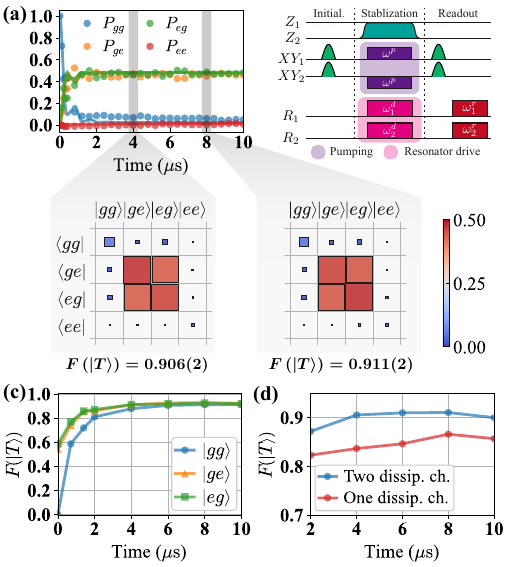}
    \caption{Stabilization of Bell state of two qubits. (a) Population dynamics of the four basis states of the two qubits as a function of time, starting from an initial state $\ket{gg}$. Markers represent experimental data and lines denote numerical simulations with master equation. Both readout resonators participate in stabilization. Two QST measurements are performed at 4 and 8 $\mu$s, with the results depicted in the two callouts. Both the size and color of the squares in the QST results represent the magnitude of the density matrices, showing that the decay into $\ket{gg}$ and the residual population in $\ket{S}$ account for most of the error. The fidelity values are calculated as described in the main text. (b) Experimental pulse sequence for state preparation, stabilization, and readout. (c) Fidelity of stabilization over time for different initial states. Once the stabilization process commences, all initial states converge towards the target Bell state. (d) Fidelity of stabilization over time when activating one or two dissipative channels (i.e., applying pulses to one or both resonators). With two channels energized, stabilization proceeds more rapidly and achieves a higher fidelity.}
    \label{fig2}
\end{figure}

Figure \ref{fig2}(a) shows one example of the stabilization process for an initial state $\ket{gg}$. In this measurement, the two qubits are tuned into resonance. After applying all drives to the qubits and resonators, the system initially undergoes a brief period where three of the four basis states are populated but soon become equally populated in the $\ket{eg}$ and $\ket{ge}$ states only. Throughout the entire period, the $\ket{ee}$ state remains almost unpopulated, as it is decoupled from the other states, as previously explained. To verify that the steady state of the stabilization is indeed the target Bell state, two QST measurements are performed at 4 and 8 $\mu$s. The fidelity is calculated as $F$ = Tr$(\bra{T}\rho_{\rm{exp}} \ket{T})$, where $\rho_{\rm{exp}}$ is constructed from QST data (\cref{fig2}(a)). $F = 90.6(2)\%$ and $91.1(2)\%$ are obtained for the two measurements without post-selection. It is noted that the time span of 10 $\mu$s in our measurements is set by the longest buffer time of the electronics, not by any intrinsic limit.

We further demonstrate that the entanglement stabilization is insensitive to the initial state. Figure \ref{fig2}(c) shows the fidelity measured at different phases of the stabilization for three different initial states. In all cases, the fidelity quickly converges and stabilizes. Due to their greater overlap with the target state, both $\ket{ge}$ and $\ket{eg}$ states approach $\ket{T}$ more rapidly than the state $\ket{gg}$. A time constant of $T_s = 0.9\ \mu$s for the stabilization process can be extracted from the data, corresponding to a stabilization rate of $\Gamma_s=1/T_s$. Other relevant rates include $\Gamma_1^{1 (2)}=1/T_1^{1 (2)}$ and $\Gamma_{\phi}=1/T_{\phi}^{\ket{S}-\ket{T}}$, which represent the rates at which the $Q_1$–$Q_2$ system is scattered away from the target state. In this case, $T_1^{1}=T_1^{2}$ = 27 $\mu$s, and $T_{\phi}^{\ket{S}-\ket{T}}$ = 18 $\mu$s are determined by the spin-echo time average of the two qubits. Using a rough estimate\cite{supplement} of $(\Gamma_s-(\Gamma_1^1+\Gamma_1^2)/2-\Gamma_{\phi})/\Gamma_s=0.917$, we arrive at a value consistent with the measured fidelities. Note that stabilization from the decoupled state $\ket{ee}$ is also possible with additional coherent Rabi driving to connect $\ket{ee}$ and $\ket{S}$\cite{supplement}.

The Raman process discussed above involves applying detuned drive(s) to the readout resonator(s), which can be done on either one or both resonators. When only one resonator is driven, the other remains in its vacuum state, resulting in the system stabilizing to a more mixed state with a slower decay rate compared to when both resonators are driven, as shown in \cref{fig2}(b). This is expected because utilizing only one dissipative channel weakens the system's ability to remove entropy. Hence, the fidelity of the target state remains unstable even after 6 $\mu$s, with its steady-state value being lower than when both dissipative channels are energized. This result highlights the flexibility of using individually controllable cavities over a common resonator for bath engineering\cite{Gourgy2015}.

\begin{figure}[htbp]
    \centering
    \includegraphics{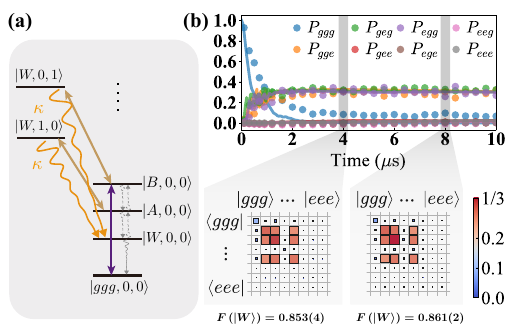}
    \caption{Stabilization of the $W$ state of three qubits. (a) Energy level diagram of the system depicting the relevant processes for stabilization. Two Raman processes utilizing different resonators engineer the paths $\ket{B} \rightarrow \ket{W}$ and $\ket{A} \rightarrow \ket{W}$. The resonant transition $\ket{ggg}\rightarrow \ket{B}$ is induced by driving $Q_2$ at a Rabi frequency $\Omega$. (b) Dynamics of the population of the eight basis states of the three qubits over time, with the system initially in $\ket{ggg}$. Two QST measurements are performed at 4 and 8 $\mu$s.}
    \label{fig3}
\end{figure}

Next, we extend our protocol to stabilize the three-qubit $W$ state. In a three-qubit system, the coupling rates are set to $J_{12}/2\pi=J_{23}/2\pi=J/2\pi=\SI{5}{MHz}$. Notice the $\ket{W}=(\ket{egg}+\ket{geg}+\ket{gge})/\sqrt{3}$ state is not an eigenstate if all qubits are in resonance. To overcome this issue, the frequency of qubit $Q_2$ (see \cref{fig1}(a)) is detuned from those of $Q_1$ and $Q_3$ by $\Delta^q_2/2\pi=J=\SI{5}{MHz}$ so that the $\ket{W}$ state, along with two other states $\ket{A}$ and $\ket{B}$, become the eigenstates in the single-excitation manifold of the system (see SM\cite{supplement} for more details about states $\ket{A}$ and $\ket{B}$).

Similar to the Bell state stabilization, dissipative channels are engineered so that the $\ket{W}$ state becomes the steady state of the driven-dissipative system with the assistance of the Raman process and rapid dissipation in the readout resonators. However, due to the more complex energy landscape of the system, two separate paths are now used to steer $\ket{A}$ and $\ket{B}$ to the $\ket{W}$ state, as illustrated in \cref{fig3}(a). Again, to counterbalance the relaxation from the target state to the ground state $\ket{ggg}$, $Q_2$ is driven at frequency $\omega^p=\omega_{\ket{B}}$ with a Rabi frequency $\Omega^p/2\pi=\SI{0.74}{MHz}$, which pumps $\ket{ggg}$ back to the single-excitation manifold. Due to the dynamical symmetry of the pumping drive, the resonant transitions to the higher excitation manifolds are forbidden\cite{supplement}. For an initial state of $\ket{ggg}$, the driven-dissipative system stabilizes after around 4 $\mu$s, with fidelity obtained via QST measurements consistently over $85\%$. The lower fidelity of the stabilized $\ket{W}$ state compared to the Bell state is expected, as there are more dephasing paths, and only one dissipative channel is engineered to counteract the effect of each path. Additionally, the probability of leakage to higher excitation states via the pumping drive is larger due to a smaller energy gap between $\ket{W}$ and $\ket{A}$ at the same coupling strength.

In conclusion, we have demonstrated a hardware-efficient scheme for generating entangled states via engineered dissipative channels. The scheme leverages the individual controllability of each qubit and its associated local readout resonator to realize precise quantum bath engineering, making it compatible with mainstream scalable architectures for superconducting quantum computation. The two-qubit Bell state and the three-qubit $W$ state have been prepared and preserved indefinitely, with state fidelities (by averaging the QST data from 4 to 10 $\mu$s with interval 2 $\mu$s) of $90.7\%$ and $86.2\%$, respectively. Further improvement of this scheme is possible through additional parity measurements\cite{LiuPRX}, optimizing parameters with larger $\kappa$ \textemdash a necessary ingredient in QEC\cite{Krinner2022}, and enhancing the coherence performance of qubits (see SM\cite{supplement}). Moreover, since the present work only uses simple time-independent drives, introducing time-dependent control techniques such as Floquet engineering\cite{Petiziol2022prl} or digital circuits\cite{xiaomi2024} could provide more flexible and effective solutions for sophisticated entanglement generation requirements.

\begin{acknowledgments}
The authors thank Youpeng Zhong, Jian Li, and Kai Luo for technical support. This work was supported by the Key-Area Research and Development Program of Guangdong Province (Grant No. 2018B030326001), the National Natural Science Foundation of China (No. 12074166 and No. 12004162), the Guangdong Provincial Key Laboratory (Grant No. 2019B121203002).
\end{acknowledgments}

\nocite{*}
\bibliography{Entanglement}

\end{document}


\title{Supplemental Material of ``Hardware-Efficient Stabilization of Entanglement via Engineered Dissipation in Superconducting Circuits"}

\input{author}
\date{\today}

\maketitle

\tableofcontents

\newpage

\section{Theoretical Framework for Bath-Engineering Protocol}

\subsection{Engineered Dissipation Theory for Eigenstates Stabilization}

In this section, we formulate the theory of the engineered dissipation protocol for entanglement stabilization used in this work. A one-dimensional array of superconducting transmon qubits with dedicated resonators is considered in this model. The corresponding Hamiltonian is given by ($\hbar=1$):

\begin{equation}
    \begin{aligned}
        H_1 &= H_q + H_r + H_{\rm{int}},\\
        H_q &= \sum_{i=1}^L (\omega^q_i b_i^\dagger b_i + \frac{\alpha_i}{2} b_i^\dagger b_i^\dagger b_ib_i) - \sum_{i=1}^{L-1} J_{i,i+1} (b_i^\dagger b_{i+1} + b_{i+1}^\dagger b_i),\\
        H_r &= \sum_{i=1}^L \omega^r_i c_i^\dagger c_i,\\
        H_{\rm{int}} &= \sum_{i=1}^L g_i(c_i^\dagger b_i+b_i^\dagger c_i).
    \end{aligned}
\end{equation}
Here, $H_q$, $H_r$, and $H_{\rm{int}}$ represent the Hamiltonians of the transmon qubits, their resonators, and the interaction between the qubits and resonators, respectively. The total number of qubits is denoted by $L$. The frequency and anharmonicity of the $i$-th qubit, $Q_i$, are denoted by $\omega^q_i$ and $\alpha_i$, respectively. The frequency of the $i$-th resonator, $R_i$, is denoted by $\omega^r_i$. The hopping strength between adjacent qubits, $Q_i$ and $Q_{i+1}$, is denoted by $J_{i,i+1}$. The coupling strength between $Q_i$ and $R_i$ is represented by $g_i$.

To engineer dissipative channels as desired for entanglement stabilization, appropriate coherent drives are applied to the dissipative resonators to enable the Raman process \cite{Murch2012,hacohen-gourgy_cooling_2015}, as discussed in the main manuscript. As all resonators are dispersively coupled to their corresponding qubits and lack direct coupling between themselves, their effects on the qubit array are individually controllable. For simplicity, consider a case where a coherent drive is applied to only one of the resonators, $R_k$. The corresponding driving term takes the following form:
\begin{equation}
    \begin{aligned}
        H_{\rm{d},k} &= \epsilon_k^d(e^{i\omega^d_k t}c_k+e^{-i\omega^d_k t}c_k^\dagger).
    \end{aligned}
\end{equation}
Here, $\epsilon_k^d$ and $\omega^d_k$ denote the amplitude and frequency of the drive. In the rotating frame defined by the unitary transformation $U=\exp(\sum_i i\omega^d_k(b_i^\dagger b_i + c_i^\dagger c_i) t)$, the total Hamiltonian of the system is given by $\Tilde{H}=UH_2U^\dagger-i U\partial_t U^\dagger$, where $H_2=H_1+H_{\rm{d},k}$. We further apply the rotating-wave approximation to eliminate the fast-varying terms in the Hamiltonian and arrive at:
\begin{equation}
    \begin{aligned}
        \Tilde{H}&=\Tilde{H}_q+\Tilde{H}_r+\Tilde{H}_{\rm{int}}+\Tilde{H}_{\rm{d},k},\\
        \Tilde{H}_q &= \sum_{i=1}^{L} (\Delta^q_i b_i^\dagger b_i + \frac{\alpha_i}{2} b_i^\dagger b_i^\dagger b_ib_i) - \sum_{i=1}^{L-1} J_{i,i+1} (b_i^\dagger b_{i+1} + b_{i+1}^\dagger b_i),\\
        \Tilde{H}_r &= \sum_{i=1}^L \Delta^r_i c_i^\dagger c_i,\\
        \Tilde{H}_{\rm{int}} &= \sum_{i=1}^L g_i(c_i^\dagger b_i+b_i^\dagger c_i),\\
        \Tilde{H}_{\rm{d},k} &= \epsilon_k^d(c_k+c_k^\dagger),
    \end{aligned}
\end{equation}
where $\Delta^q_i=\omega^q_i-\omega^d_k$ and $\Delta^r_i=\omega^r_i-\omega_k^d$ denote the detunings between the frequencies of the qubits and resonators and the driving frequency, respectively.

To elucidate the effective Hamiltonian in the Raman process discussed in the main manuscript, we adopt an analysis similar to that in Ref.\cite{hacohen-gourgy_cooling_2015}. Initially, the anharmonicity of the qubits and the driving terms in $\Tilde{H}$ are disregarded, as their contributions can be considered perturbative given that $\omega_i^r,\omega_i^q \gg \alpha_i,\epsilon_k^d$ is satisfied in our experiment. With this simplification, the system can be viewed as a set of coupled harmonic oscillators. To determine the normal modes of the system, $H_0$ is expressed in matrix form as:
\begin{equation}
    \begin{aligned}
        H_0 &= \sum_{i=1}^{L} \Delta^q_i b_i^\dagger b_i + \sum_{i=1}^{L} \Delta^r_i c_i^\dagger c_i - \sum_{i=1}^{L} J_{i,i+1} (b_i^\dagger b_{i+1} + b_{i+1}^\dagger b_i) + \sum_{i=1}^{L} g_i(c_i^\dagger b_i+b_i^\dagger c_i)\\
        &=\bm{\textbf{v}}^\dagger\mathcal{H}_0\bm{\textbf{v}}.
    \end{aligned}
\end{equation}
where $\textbf{v}$ is a $2L$-component vector $(b_1,\cdots b_L,c_1,\cdots c_L)$, and the $2L\times 2L$ matrix $\mathcal{H}_0$ encompasses all the frequencies of the qubits and resonators and their couplings. It is important to note that $H_0$ contains only quadratic terms. Using standard linear algebra, $H_0$ can be diagonalized to obtain the normal modes of the coupled system. Next, a matrix $M$ is constructed from the orthonormal eigenvectors of $\mathcal{H}_0$, satisfying $M^{-1}=M^{T}$. The matrix $M^{-1}\mathcal{H}_0 M$ is diagonal, with the corresponding eigenmodes given by $\bm{\textbf{W}}=M^{-1}\bm{\textbf{v}}$. In the new basis vector $\bm{\textbf{W}}=(B_1,\cdots,B_L,C_1,\cdots,C_L)$, $H_0$ becomes a sum of uncoupled harmonic oscillators:
\begin{equation}
    H_0 = \sum_{i=1}^L  \lambda^r_i C_i^\dagger C_i+\lambda^q_i B_i^\dagger B_i,
\end{equation}
where $\lambda_i^{q,r}$ denote the respective normal-mode frequencies.

We would like to make some comments before further discussion. First, the transmon qubits and their readout resonators are coupled in the dispersive regime ($g_i \ll \left| \omega_i^q-\omega_i^r\right|$), therefore the dressed modes $C_i$ consist a large portion of the resonator modes, with small contributions from the qubit modes. Second, the resonator-like modes $C_i$ mainly contain the bare mode $c_i$ due to negligible couplings between the resonators and off-resonant conditions for all resonators. This fact enables the activation of specific Raman processes for entanglement stabilization. Third, the $B_i$ modes are primarily linear combinations of transmon excitations $(b_1,\cdots b_L)$ with small components from the resonators $(c_1,\cdots c_L)$.

We now focus on the anharmonicity of the transmon qubits neglected before. In terms of the new normal modes, the anharmonic terms can be expressed as
\begin{equation}
    \begin{aligned}
       \sum_{i=1}^L \frac{\alpha_i}{2} b_i^\dagger b_i^\dagger b_ib_i=\frac{1}{2}\sum_{i=1}^L \alpha_i
       \left[\sum_{s=1}^{L}(M_{i(L+s)}C_{s} + M_{is}B_{s})\right]^\dagger
       \left[\sum_{u=1}^{L}(M_{i(L+u)}C_{u} + M_{iu}B_{u})\right]^\dagger\\
       \left[\sum_{l=1}^{L}(M_{i(L+l)}C_{l} + M_{il}B_{l})\right]
       \left[\sum_{p=1}^{L}(M_{i(L+p)}C_{p} + M_{ip}B_{p})\right],
    \end{aligned}
\end{equation}
Given the large detunings between the qubits and resonators, terms that do not conserve the excitation number of the qubits or resonators can be neglected under the rotating wave approximation. The anharmonic terms can then be categorized into three groups \cite{hacohen-gourgy_cooling_2015}:

$\bullet$ self-Kerr terms of qubit-like operators, $\sum_{s,u,l,p=1}^{L}\frac{1}{2}\mu_{sulp}B_s^\dagger B_u^\dagger B_l B_p$, where $\mu_{sulp}$ is given by:
\begin{equation}
    \mu_{sulp} = \sum_{i=1}^{L}\alpha_iM_{is}M_{iu}M_{il}M_{ip};
\end{equation}

$\bullet$ self-Kerr terms of resonator-like operators, $\sum_{s,u,l,p=1}^{L}\frac{1}{2}\xi_{sulp}C_s^\dagger C_u^\dagger C_l C_p$, where $\xi_{sulp}$ is given by:
\begin{equation}
    \xi_{sulp} = \sum_{i=1}^{L}\alpha_iM_{i(L+s)}M_{i(L+u)}M_{i(L+l)}M_{i(L+p)};
\end{equation}

$\bullet$ cross-Kerr terms between qubit-like and resonator-like operators, $\sum_{s,u,l,p=1}^{L}2\eta_{sulp}C_s^\dagger C_u B_l^\dagger B_p$, where $\eta_{sulp}$ is given by:
\begin{equation}
    \eta_{sulp} = \sum_{i=1}^{L}\alpha_iM_{i(L+s)}M_{i(L+u)}M_{il}M_{ip}.
\end{equation}
By the definition of $M$ and $\bm{\textbf{W}}=M^{-1}\bm{\textbf{v}}$, $M_{ij}$ represents the overlap between the bare mode $b_i$ and the dressed mode $B_j$ ($C_{j-L}$) for $1\le j\le L$ ($L+1\le j \le 2L$). Since qubit-like operators $B_{j}$ are primarily linear combinations of bare qubit modes $b_{i}$, the following approximation holds:
\begin{equation}
M_{i(L+j)} \ll M_{ij}, (1\le i,j\le L).\label{eq:m_rel}
\end{equation}
It follows that $\xi_{sulp}\ll\eta_{sulp}\ll\mu_{sulp}$. While the qubit-like self-Kerr terms are the largest, these operators are not important in our experiment as we focus on the subspace $\{\ket{g},\ket{e}\}$ for all qubits; thus, both types of self-Kerr terms can be neglected. For the cross-Kerr terms, $C_s^\dagger C_u (s\ne u)$ are fast-varying compared to $C_s^\dagger C_s$ and can also be dropped. Consequently, only cross-Kerr terms in the form $\sum_{n,l,p=1}^{L}2\eta_{nnlp}C_n^\dagger C_n B_l^\dagger B_p$ remain. Consider a particular term 
\begin{equation}
    H_{\mathrm{X-Kerr,k}} = \sum_{l,p=1}^{L}2\eta_{kklp}C_k^\dagger C_k B_l^\dagger B_p,
\end{equation}
where $\eta_{kklp} = \sum_{i=1}^{L}\alpha_iM_{i(L+k)}^2M_{il}M_{ip}$. This term describes the contribution of a coherent drive applied to the $k$-th resonator $R_k$. In the following, we show how $H_{\mathrm{X-Kerr,k}}$ induces the Raman process in our protocol.

An intuitive understanding of the role of $H_{\mathrm{X-Kerr,k}}$ can be achieved by following the approach discussed in Refs. \cite{hacohen-gourgy_cooling_2015, Murch2012}. The key point is that, under a classical microwave driving, the dissipative resonator $R_k$ is displaced, and its operator $C_k$ can be expressed as the sum of a classical drive $\Bar{C}_k=\left<C_k\right>$ and a quantum correction $D_k$, i.e. $C_k=\Bar{C}_k+D_k$, where $\Bar{n}_k=|\Bar{C}_k|^2$ is the average photon number in the resonator. Therefore we have
\begin{equation}
    \begin{aligned}
        \Tilde{H}_{\mathrm{X-Kerr,k}} &= \sum_{l,p=1}^{L} 2\eta_{kklp}(\Bar{C}_k+D_k)^\dagger(\Bar{C}_k+D_k)B_l^\dagger B_p\\
        &= \sum_{l,p=1}^{L} 2\eta_{kklp}(\Bar{C}_k^*\Bar{C}_k+\Bar{C}_k^*D_k+\Bar{C}_kD_k^\dagger+D_k^\dagger D_k)B_l^\dagger B_p.
    \end{aligned}
\end{equation}
Using the condition in \cref{eq:m_rel}, we have
\begin{equation}
    \begin{aligned}
        b_i^\dagger b_i = (\sum_{l=1}^{L}M_{i(L+l)}C_l + M_{il}B_l)(\sum_{p=1}^{L}M_{i(L+p)}C_p + M_{ip}B_p)
        \approx \sum_{l,p=1}^{L} M_{il}M_{ip}B_l^\dagger B_p,
    \end{aligned}
\end{equation}
then the first term in $\Tilde{H}_{\mathrm{X-Kerr,k}}$ can be rewritten as:
\begin{equation}
    \begin{aligned}
        \sum_{l,p=1}^{L} 2\eta_{kklp}\Bar{C}_k^*\Bar{C}_k B_l^\dagger B_p
        &= \sum_{i,l,p=1}^{L} 2\Bar{C}_k^*\Bar{C}_k\alpha_i M_{i(L+k)}^2 M_{il}M_{ip} B_l^\dagger B_p\\
        &= \sum_{i=1}^{L} 2\Bar{C}_k^*\Bar{C}_k\alpha_i M_{i(L+k)}^2 \sum_{l,p=1}^{L} M_{il}M_{ip} B_l^\dagger B_p\\
        &= \sum_{i=1}^{L} 2\Bar{C}_k^*\Bar{C}_k\alpha_iM_{i(L+k)}^2 b_i^\dagger b_i\\
        &= \sum_{i=1}^{L} 2\Bar{n}_k\alpha_iM_{i(L+k)}^2 b_i^\dagger b_i.
    \end{aligned}
\end{equation}
This represents the frequency shift of the qubits (almost only $Q_k$) induced by the average photon number $\Bar{n}_k$ in the resonator $R_k$, which can be compensated by fine-tuning the qubit frequency in our experiment.

The two terms $D_k \Bar{C}_k^*$ and $D_k^\dagger D_k$ in $\Tilde{H}_{\mathrm{X-Kerr,k}}$ can be neglected if the loss rate of the resonator $R_k$ is significantly higher than the dynamics of $\Tilde{H}_{\mathrm{X-Kerr,k}}$. The remaining term in $\Tilde{H}_{\mathrm{X-Kerr,k}}$, referred to as $V_{\mathrm{cool}}$ (cooling operator) below, is the most relevant term for the Raman process and is given by
\begin{equation}
    V_{\mathrm{cool}} = \sum_{l,p=1}^{L}2\eta_{kklp}\Bar{C}_kD^\dagger_kB_l^\dagger B_p
    = \sum_{l,p=1}^{L} d^k_{lp}D^\dagger_kB_l^\dagger B_p,
\end{equation}
where $d^k_{lp}$ is defined as 
\begin{equation}
    d_{lp}^k = 2\Bar{C}_k\eta_{kklp} = \sum_{i=1}^L2\Bar{C}_k\cdot\alpha_i M_{i(L+k)}^2\cdot M_{il}M_{ip}.
\end{equation}

When the resonator $R_k$ is driven by a red-detuned drive $\omega_k^d$, the cooling operator $V_{\mathrm{cool}}$ scatters an excited eigenmode ($B_p$) into another excited eigenmode ($B_l^\dagger$) with lower energy by creating a photon ($D_k^\dagger$) with a higher frequency than the incoming one. Specifically, an inelastic scattering occurs in which a driving photon is converted into one at the resonator frequency when the condition $\Delta_k^r=\omega_k^r-\omega_k^d=\lambda^q_p-\lambda_l^q$ is met. Consequently, some excess energy is transferred from the qubit array to the resonator, followed by dissipation into the environment via the fast decay of the resonator. Ultimately, the qubit array is left in the excited eigenmode $B_l$.

Since the resonator $R_k$ only couples to the qubit $Q_k$, in the dispersive regime $|\Delta_k^{rq}|=|\omega_k^r-\omega_k^q|\gg g_k$, the parameter $M_{i(L+k)}$ in the cross-Kerr terms can be approximated as $M_{i(L+k)} \approx g_k \delta_{ik}/\Delta_k^{rq}$, where $\delta_{ik}=1$ for $i=k$ and 0 otherwise. Therefore, we have
\begin{equation}
    d^k_{lp}\approx\sum_{i=1}^{L}2\Bar{C}_k\cdot\alpha_i\left(\frac{g_{k}\delta_{ik}}{\Delta^{rq}_k}\right)^2\cdot M_{il}M_{ip}=2\Bar{C}_k\cdot\alpha_k\left(\frac{g_{k}}{\Delta^{rq}_k}\right)^2\cdot M_{kl}M_{kp}.\label{eq:trans_rate}
\end{equation}
In our experiment, $|\Delta^{rq}_k|\gg|\alpha_k|$, so the dispersive shift $\chi_{kk}$ between $R_k$ and $Q_k$  is approximated by $\chi_{kk}\approx{g_{k}^2\alpha_k}/{\Delta^{rq}_k(\Delta^{rq}_k-\alpha_k)} \approx \alpha_k({g_{k}}/{\Delta^{rq}_k})^2$\cite{Koch2007}. Additionally, with the average photon number $\Bar{n}_k=|\Bar{C}_k|^2$ in the resonator $R_k$, $d^k_{lp}$ can be further expressed as follows:
\begin{equation}
    \begin{aligned}
        d^k_{lp} \approx 2\sqrt{\Bar{n}_k}\cdot\chi_{kk}\cdot M_{kl}M_{kp}.
    \end{aligned}
\end{equation}

While $\sum_{i=1}^{L}M_{il}M_{ip} \approx 0$ for $l\neq p$, $M_{il}M_{ip}\ne 0$ generally holds as it represents the wavefunction overlap between the eigenmodes $B_l$ and $B_p$ at qubit $Q_i$. The distinction between our setup, where each qubit has a dedicated readout resonator, and configurations where qubits are coupled to a common resonator\cite{hacohen-gourgy_cooling_2015}, should be emphasized. In the latter case, assuming all qubits have similar coupling to the resonator and similar anharmonicity, the transition matrix of \cref{eq:trans_rate} becomes $d_{lp}\approx 2\Bar{C}\alpha\left(\frac{g}{\Delta^{rq}}\right)^2\sum_{i=1}^{L}M_{il}M_{ip} \approx0$ due to the orthogonality of eigenstates. Furthermore, in our setup, different readout resonators can be utilized for different Raman processes to enhance the overall speed, as demonstrated in the main manuscript, clearly showcasing the advantages of dedicated resonators over a shared resonator.

\subsection{Engineered Dissipative Rate with Raman Process}

To evaluate the engineered dissipative rate $\Gamma_{p\rightarrow l}^k$ from eigenmode $B_p$ to $B_l$ when the resonator $R_k$ is driven, Fermi's golden rule is applied as follows:
\begin{equation}
    \Gamma_{p\rightarrow l}^k = \left|\matrixel{\Psi_l}{V_{\mathrm{cool}}}{\Psi_p}\right|^2 S_k(\Delta^r_k)=4\Bar{n}_k|\chi_{kk}M_{kl}M_{kp}|^2\frac{\kappa_k}{(\lambda^q_p-\lambda^q_l-\Delta^r_k)^2+(\kappa_k/2)^2}.
\end{equation}
where $\left|\matrixel{\Psi_l}{V_{\mathrm{cool}}}{\Psi_p}\right|$ is the transition matrix element associated with the Raman process, and the Lorentzian shape $S_k(\Delta^r_k)=\frac{\kappa_k}{(\lambda^q_p-\lambda^q_l-\Delta^r_k)^2+(\kappa_k/2)^2}$ represents the spectral density of the photon number fluctuations in resonator $R_k$\cite{hacohen-gourgy_cooling_2015}. The optimal condition for fast $\Gamma_{p\rightarrow l}^k=16\Bar{n}_k|\chi_{kk}M_{kp}M_{kl}|^2/\kappa_k$ is achieved when $\Delta^r_k=\omega^r_k-\omega^d_k=\lambda^q_p-\lambda^q_l$. Under this condition, the undesired reverse process $\Gamma_{l\rightarrow p}^k$ is  
\begin{equation}
    \Gamma_{l\rightarrow p}^k = 4\Bar{n}_k|\chi_{kk}M_{kp}M_{kl}|^2\frac{\kappa_k}{4(\lambda^q_p-\lambda^q_l)^2+(\kappa_k/2)^2}.
\end{equation}
Consequently, the ratio between the two rates is  
\begin{equation}
\Gamma_{p\rightarrow l}^k/\Gamma_{l\rightarrow p}^k= \frac{4(\lambda^q_p-\lambda^q_l)^2+(\kappa_k/2)^2}{(\kappa_k/2)^2} \approx 16\left(\frac{\lambda^q_p-\lambda^q_l}{\kappa_k}\right)^2+1 \gg 1,  \text{when } |\lambda^q_p-\lambda^q_l| \gg \kappa_k.
\end{equation}
The condition $|\lambda^q_p-\lambda^q_l| \gg \kappa_k$ is easily satisfied in our experiment, thus an approximate unidirectional dissipative process from eigenmode $B_p$ to $B_l$ is effectively
engineered with the assistance of the Raman process and the dissipative resonator.

\subsection{Pumping Drive with Dynamical Symmetry}

To further mitigate the leakage from the single-excitation manifold into the ground state due to the finite $T_1$ of qubits, a coherent microwave drive must be applied to the qubit array to pump the ground state back to the single-excitation manifold. However, transitions from the single-excitation manifold to the higher-excitation manifold could occur if the pumping drive is inappropriate. To this end, a pumping drive with dynamical symmetry is applied to suppress undesired transitions in this work, as discussed below.

Consider two eigenstates of the qubit array in the single-excitation manifold, $B_l^\dagger \ket{G}$ and $B_p^\dagger \ket{G}$, where $B_l \approx \sum_{i=1}^{L} M_{il} b_i$ and $B_p \approx \sum_{i=1}^{L} M_{ip} b_i$, and $\ket{G}$ represents the ground state. The eigenfrequencies of $B_l^\dagger \ket{G}$ and $B_p^\dagger \ket{G}$ are $\lambda^q_l$ and $\lambda^q_p$, respectively. The pumping drive applied to all qubits with the same frequency $\omega^p$ can be expressed as $H_{\mathrm{pump}} = \sum_{i=1}^{L} \Omega_i^p b_i e^{i\omega^p t}+h.c.$, where $\Omega^p_i$ is the complex amplitude of the drive applied to qubit $Q_i$. Suppose we only want to pump the ground state $\ket{G}$ back to the eigenstate $B_p^\dagger \ket{G}$. Then the parameters of the pumping drive should satisfy $\omega^p=\lambda^q_p$ and $\Omega_i^p/M_{ip}\equiv \mathrm{const.}$ for all $i$. In other words, $H_{\mathrm{pump}} \propto (B_p+B_p^{\dagger})$ in the rotating frame at frequency $\omega^p=\lambda^q_p$. One can verify that under the above conditions, $\bra{G}B_l H_{\mathrm{pump}}\ket{G}=0$ for $l\neq p$, meaning the ground state only transitions to the eigenstate $B_p^\dagger\ket{G}$ under the pumping drive $H_{\mathrm{pump}}$.

\begin{figure}[htbp]
    \centering
    \includegraphics{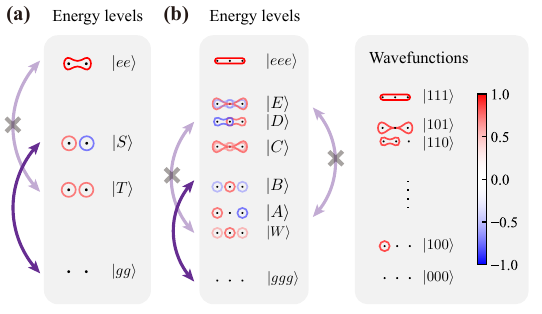}
    \caption{Eigenstates of the two-qubit (a) and three-qubit (b) arrays employed in this experiment. Different wavefunction shapes represent the eigenstates of the qubit arrays. Color indicates the amplitude of the wavefunctions. In the Bell ($W$) state experiment, the pumping drives selectively excite the transition between the ground state $\ket{gg}$ ($\ket{ggg}$) and $\ket{S}$ ($\ket{B}$) while inhibiting the transition $\ket{T}\leftrightarrow\ket{ee}$ ($\ket{W}\leftrightarrow\ket{D},\ket{A}\leftrightarrow\ket{E}$) due to the dynamical symmetry of the pumping drive. Meanwhile, the other possible transitions are suppressed by a far-detuning condition.}
    \label{fig:spec_sym}
\end{figure}

For Bell state stabilization, two qubits are tuned to resonance with a coupling strength $J$. The eigenstates in the single-excitation manifold are $\ket{T}=(\ket{ge}+\ket{eg})/\sqrt{2}$ and $\ket{S}=(\ket{ge}-\ket{eg})/\sqrt{2}$, with the ground state $\ket{G}=\ket{gg}$, as shown in \cref{fig:spec_sym}(a). In this case, the pumping drive with dynamical symmetry is chosen as $H_{\mathrm{pump}}=\Omega^p(b_1-b_2) + h.c.$ ($\Omega^p_1=-\Omega^p_2=\Omega^p$) and $\omega^p=\omega_{\ket{S}}-\omega_{\ket{gg}}$ to pump $\ket{gg}$ into $\ket{S}$, where $\Omega^p \ll J$. Combined with the engineered dissipative channel $\Gamma_{\ket{S}\rightarrow \ket{T}}$ through the Raman process, the population in state $\ket{S}$ is transferred to $\ket{T}$. It is evident that both $\bra{gg}H_{\mathrm{pump}} \ket{T}=0$ and $\bra{ee}H_{\mathrm{pump}} \ket{T}=0$ even though $\omega_{\ket{ee}}-\omega_{\ket{T}}=\omega_{\ket{S}}-\omega_{\ket{gg}}$. This is important for our Bell state stabilization as the target state $\ket{T}$ is decoupled from the pumping drive. Furthermore, the transition $\ket{S}\leftrightarrow \ket{ee} $ is highly suppressed due to its far-detuning $\Omega^p \ll 2J$.

The eigenstates of a three-qubit array for $W$ state stabilization are shown in \cref{fig:spec_sym}(b). The ground state is $\ket{G}=\ket{ggg}$. The detailed structure of the eigen-spectrum is given in \cref{eigenstate_three}. While the pumping drive $H_{\mathrm{pump}} = \sum_{i=1}^{3} \Omega^p_i b_i e^{i\omega^p t}+h.c.$ could be constructed with $\omega^p=\omega_{\ket{B}}-\omega_{\ket{G}}$ and $\Omega^p_1:\Omega^p_2:\Omega^p_3=1:-2:1$  for $\ket{G}\leftrightarrow \ket{B}$, given $\ket{B}=(\ket{egg}-2\ket{geg}+\ket{gge})/\sqrt{6}$, in our experiment, we only pump the middle qubit $Q_2$ with $\Omega^p_2=\Omega^p$ and set $\Omega^p_1=\Omega^p_3=0$ for simplicity. Such simplified pumping is still feasible thanks to the large energy gap $\omega_{\ket{B}}-\omega_{\ket{W}}=3J$ compared to the pumping amplitude $\Omega^p$ ($\Omega^p \ll 3J$) and the higher weight of $Q_2$ in $\ket{B}$ than in $\ket{W}$; thus the coherent transition $\ket{G}\leftrightarrow \ket{W}$ should be negligible. Besides, $\ket{A}=(\ket{egg}-\ket{gge})/\sqrt{2}$ is decoupled from the pumping drive as the weight of $Q_2$ in this state is zero. The more important observation is that, even though $\omega_{\ket{B}}-\omega_{\ket{G}}=\omega_{\ket{D}}-\omega_{\ket{W}}=\omega_{\ket{E}}-\omega_{\ket{A}}$, the symmetry of the pumping drive ensures that $\bra{D}H_{\mathrm{pump}}\ket{W}=\bra{E}H_{\mathrm{pump}}\ket{A}=0$. Again, all possible coherent transitions induced by the pumping drive are highly suppressed under the constraint $\Omega^p \ll J$. Overall, the coherent transition between the single-excitation manifold and the rest of the system's subspace is dominated by the resonant process $\ket{B}\leftrightarrow \ket{G}$, while others are either prohibited by the symmetry constraint or highly suppressed due to the far-detuning condition.

\subsection{Master Equation for Stabilization Process}  

The total Hamiltonian of the system in our experiment, with near-resonant drives applied to both qubits and resonators for the Raman process and pumping, is given by
\begin{equation}
    \begin{aligned}
    &H = H_q + H_r + H_{\mathrm{int}}
    + H_{\mathrm{drive}}
    + H_{\mathrm{pump}},\\
    &H_{\mathrm{drive}} = \sum_{i=1}^{L} H_{\mathrm{d},i} = \sum_{i=1}^{L} \epsilon^d_k (e^{i\omega_i^d t} c_k + h.c.),\\
    &H_{\mathrm{pump}} = \sum_{i=1}^{L} \Omega^p_i(e^{i\omega^p t} b_i + h.c.).
    \end{aligned}
\end{equation}
Taking into account the dissipation in the resonators and the decoherence of qubits, the dynamics of the density matrix of the entire system is governed by the Lindblad master equation as follows:
\begin{equation}
    \dot{\rho} = -i[H,\rho]
    +\sum_{i=1}^{L}\kappa_i\mathcal{D}(c_i)\rho
    +\sum_{i=1}^{L}\Gamma_1^i\mathcal{D}(b_i)\rho
    +\sum_{i=1}^{L}\Gamma_\phi^i\mathcal{D}(b_i^\dagger b_i)\rho.
\end{equation}
where $\mathcal{D}(L)\rho=(2L\rho L^\dagger - L^\dagger L\rho - \rho L^\dagger L)/2$ represents the dissipator for the collapse operator $L$. $\Gamma^i_1$ and $\Gamma^i_\phi$ denote the rates of the $T_1$ and $T_2$ processes of qubit $Q_i$ (see \cref{tab:device_parameter} for detailed values). The steady state of the driven-dissipative system is obtained by solving the equation at $\dot{\rho}=0$. By further tracing out the degrees of freedom of the resonators, one obtains the reduced density matrices describing the entangled states of the qubits.

\section{Additional Information of Experimental Techniques and Data Analysis}

\subsection{Experimental Setup and Device Parameters}

\begin{figure}[htbp]
    \centering
    \includegraphics{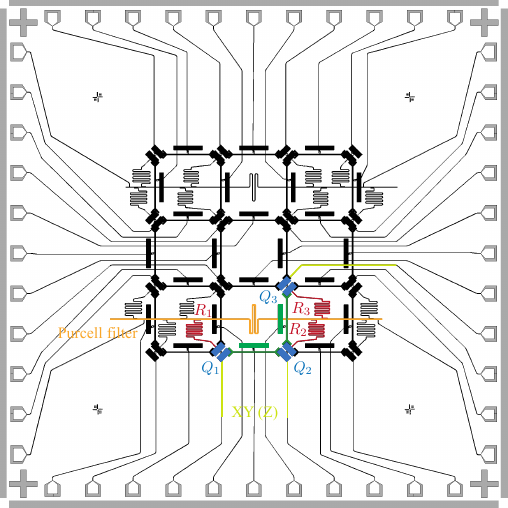}
    \caption{Superconducting quantum chip used in the experiment. It consists of a $4\times4$ array of transmon qubits and their peripheral components for control and readout. The three qubits colored blue and their resonators colored red are used in this work. Neighboring qubits are connected by tunable couplers colored green. The yellow and orange features indicate the XY(Z) control lines and the Purcell filter, respectively.}
    \label{fig:sample}
\end{figure}

The superconducting quantum chip used in our experiment comprises a $4 \times 4$ array of tunable transmon qubits in a flip-chip package, as depicted in \cref{fig:sample}. Tunable couplers\cite{yan2018} are used to control the coupling between qubits. Each qubit is capacitively coupled to a dedicated readout resonator. Two Purcell filters are designed for fast readout while protecting the qubits from decay. Each filter is connected to eight readout resonators. Key parameters are listed in \cref{tab:device_parameter}. 

\begin{table}[htbp]
    \centering
    \tabcolsep=7mm
    \renewcommand{\arraystretch}{1.2}
    \caption{Device parameters of the qubits and resonators used in our experiment.}
    \label{tab:device_parameter}
    \begin{tabular}[c]{cccc}
        \toprule
        Parameters & $Q_1$ & $Q_2$ & $Q_3$\\
        \midrule
        Qubit frequency $\omega^{q}_i/2\pi$ (GHz) & 4.202 & 4.430 & 4.179\\
        Qubit anharmonicity $\alpha_i/2\pi$ (MHz) & -197 & -189 & -199\\
        Readout resonator frequency $\omega^{r}_i$ (GHz) & 6.481 & 6.604 & 6.517\\
        Readout resonator linewidth $\kappa_i/2\pi$ (MHz) & 1.1 & 0.87 & 0.88\\
        Dispersive shift $\chi_i/2\pi$ (MHz) & -0.75 & -0.90 & -0.85\\
        Relaxtion $T_1^i$ (\si{\us}) & 27 & 27 & 27\\
        Ramsey $T_{2r}^i$ (\si{\us}) & 5.4 & 15 & 4.9\\
        Spin echo $T_{2e}^i$ (\si{\us}) & 14 & 28 & 11\\
        Working frequency (Bell state exp.) $\omega_{\text{Bell}}/2\pi$ (\si{GHz}) & 4.202 & 4.202 & \\
        Working frequency (W state exp.) $\omega_{\text{W}}/2\pi$ (\si{GHz}) & 4.179 & 4.179 & 4.179\\
        \bottomrule
    \end{tabular}
\end{table}

\subsection{Error Mitigation and 
Parameter Calibration}

\subsubsection{Idling Setup and Readout Error Mitigation}

In our experiment, all qubits, when not performing logic operations, are set to idle at their sweet points to suppress the effect of flux noise. Additionally, to reduce the side effects of residual ZZ coupling between qubits during idling or single-qubit gate operations, we bias the tunable couplers near their turn-off points \cite{yan2018}, where the ZZ coupling is negligible.

\begin{figure}[htbp]
    \centering
    \includegraphics{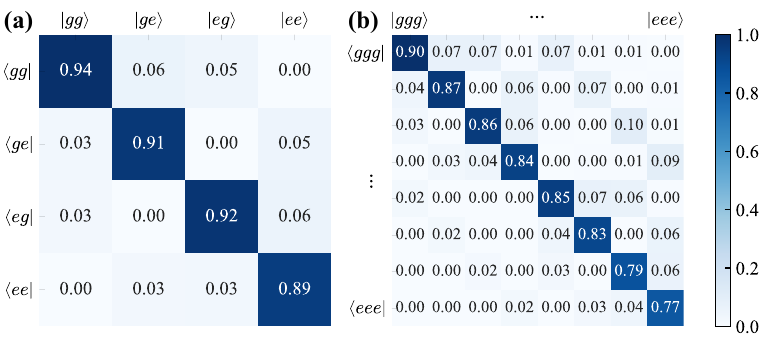}
    \caption{Matrices for readout error mitigation for the (a) two-qubit and (b) three-qubit cases. The numbers in each column represent the measured populations in different base states (listed on left of the matrices) of the prepared initial states (listed on top of the matrices).}
    \label{fig:supp_readout_matrix}
\end{figure}

Nonideal dispersive measurements\cite{Koch2007} induce readout errors in our experiments. To address this issue, we follow a commonly used protocol to generate a readout matrix $\mathcal{M}$ for correcting the readout results. $\mathcal{M}$ is obtained by preparing qubits in different states of a set of computational bases successively and measuring these states under the same bases. Two examples for the two-qubit and three-qubit cases are shown in \cref{fig:supp_readout_matrix}. The experimental readout results are then multiplied by $\mathcal{M}^{-1}$ to recover the correct populations.

\subsubsection{Correction of Z Pulse Distortion}

\begin{figure}[htbp]
    \centering
    \includegraphics[width=8.6cm]{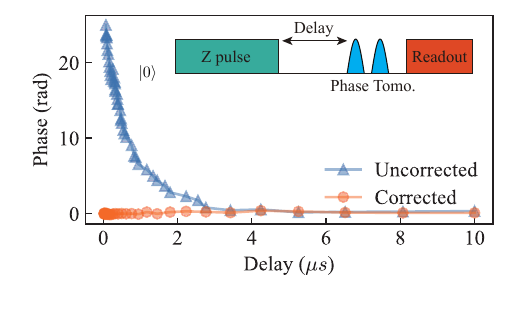}
    \caption{Residual phase after a long rectangular Z pulse, with and without predistortion correction. The inset shows the pulse sequence for the measurement.}
    \label{fig:predistortion}
\end{figure}

In our experiment, all qubits must be precisely tuned to their working frequencies. This goal is achieved by individually adjusting the flux threading the SQUID loop in each qubit via a pulse sent through its Z-control line. However, this Z-control signal may become distorted on its way from electronics at room temperature to the qubits. Such distortion must be carefully calibrated and corrected to perform high-precision experiments. For this purpose, we measure the offset of the uncorrected Z pulse relative to the ideal Z pulse using the method described in \cite{Zhiguang2019}. The inset of 
\cref{fig:predistortion} shows the measurement sequence. A Z pulse is applied to the qubit, followed by a Ramsey-like experiment to measure the phase accumulated between two $\pi/2$ pulses. In the ideal case without distortion, this phase should be zero. In reality, distortion causes a frequency change in the qubit, leading to a nonzero phase whose value depends on the amplitude of the distortion and the delay time. We first perform the measurement for the uncorrected Z pulse and then use the results to design a pre-distorted Z pulse. For a properly pre-distorted pulse, repeating the above measurement yields nearly zero phases at all delays, as shown in \cref{fig:predistortion}. At this stage, an effective correction of the Z pulse distortion is achieved.

\subsubsection{Eigenmodes of Single-excitation Manifold}

For Bell state stabilization, to ensure that the Bell states are the eigenstates of qubits $Q_1$ and $Q_2$, the two qubits are tuned to resonance. The eigenstates in the $n$-excitation manifold are as follows:
\begin{equation}
    \begin{aligned}
        n=0,\quad & \left\{\ket{gg}\right\};\\
        n=1,\quad & \left\{\ket{T}=\frac{\ket{eg}+\ket{ge}}{\sqrt{2}},
        \ket{S}=\frac{\ket{eg}-\ket{ge}}{\sqrt{2}}\right\};\\
        n=2,\quad & \left\{\ket{ee}\right\}.
    \end{aligned}
\end{equation}
One of the Bell states, $\ket{T}$ or $\ket{S}$, is chosen as the target state for stabilization.

For the three-qubit case, note that the $W$ state is not an eigenstate of the system when the three qubits are tuned to resonance with the same coupling $J$. Therefore, the parameters need to be properly modified so that the $W$ state becomes an eigenstate of the system. This goal is achieved by detuning the middle qubit $Q_2$ by $\Delta_2=J$ from $Q_1$ and $Q_3$. The eigenstates in the $n$-excitation manifold are then as follows:
\begin{equation}
    \begin{aligned}
        n=0,\quad & \{\ket{ggg}\};\\
        n=1,\quad & \{\ket{W}=\frac{\ket{gge}+\ket{geg}+\ket{egg}}{\sqrt{3}},
        \ket{A}=\frac{\ket{gge}-\ket{egg}}{\sqrt{2}},
        \ket{B}=\frac{\ket{gge}-2\ket{geg}+\ket{egg}}{\sqrt{6}}\};\\
        n=2,\quad & \{\ket{C}=\frac{\ket{eeg}+2\ket{ege}+\ket{gee}}{\sqrt{6}},
        \ket{D}=\frac{\ket{eeg}-\ket{gee}}{\sqrt{2}},
        \ket{E}=\frac{\ket{eeg}-\ket{ege}+\ket{gee}}{\sqrt{3}}\};\\
        n=3,\quad & \{\ket{eee}\}.
    \end{aligned}\label{eigenstate_three}
\end{equation}

\begin{figure}[htbp]
    \centering
    \includegraphics{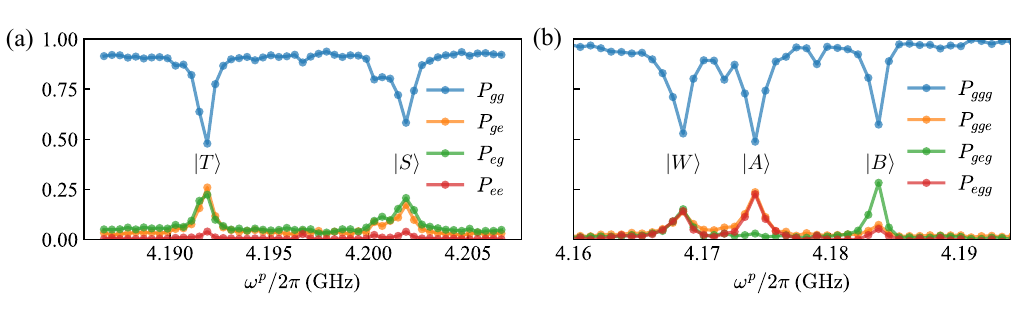}
    \caption{Energy spectra of the two-qubit (a) and three-qubit (b) systems used in this work.}
    \label{fig:energy_spectrum}
\end{figure}

Next, we experimentally manifest the eigenstates in the $1$-excitation manifold using a spectroscopy measurement. For this purpose, we apply a long (a few microseconds) square XY pulse to one of the qubits and measure the populations of different states as a function of pulse frequency. The results for two and three qubits are shown in \cref{fig:energy_spectrum}. Important information can be revealed by this simple measurement. First, the system is excited at specific frequencies corresponding to its eigenenergies. Second, the population distribution among different basis states helps identify the eigenstates. For example, the two eigenmodes observed in \cref{fig:energy_spectrum}(a) have equal weight on the bases $\ket{ge}$ and $\ket{eg}$, consistent with the two Bell states $\ket{T}$ and $\ket{S}$. Similarly, in \cref{fig:energy_spectrum}(b), three eigenmodes can be identified according to their relative weights on various basis states. The population distribution of different eigenstates matching that of the desired states indicates that our experimental parameters are well-tuned. Third, we also note that a small but nonzero population arises on the basis $\ket{ee}$ in the two-qubit case. This unwanted excitation happens because the XY drive pulse is applied only to one qubit, and can be further suppressed by driving both qubits simultaneously with appropriate amplitude and phase.

\subsubsection{Parameters of Pumping Drive}

To mitigate relaxation to the ground state due to the $T_1$ process, we apply a symmetry-selective drive that couples the ground state to an intermediate state while decoupling the system from higher excitation manifolds. For Bell state stabilization, the drive consists of simultaneously driving both qubits with the same frequency and amplitude but in opposite phases. For $W$ state stabilization, only the middle qubit is pumped. In our experiment, an iSWAP-like operation \cite{Schwartz2016} is used to synchronize the phases. The parameters of the drives used in our experiment are listed in \cref{tab:pump_parameter}.

\begin{table}[htbp]
    \centering
    \tabcolsep=7mm
    \renewcommand{\arraystretch}{1.2}
    \caption{Parameters of the pumping drives applied to qubits in our experiments.}
    \label{tab:pump_parameter}
    \begin{tabular}[c]{cccc}
        \toprule
        Bell state exp. & $Q_1$ & $Q_2$ & $Q_3$\\
        \midrule
        Pumping Amplitude $\Omega^p_i/2\pi$ (\si{MHz}) & 0.53 & -0.53 & \\
        Pumping Frequency $\omega^p/2\pi$ (\si{MHz}) & $\omega_{\ket{S}}$ & $\omega_{\ket{S}}$ \\
        \midrule
        W state exp. & $Q_1$ & $Q_2$ & $Q_3$\\
        \midrule
        Pumping Amplitude $\Omega^p_i/2\pi$ & 0 & 0.74 & 0\\
        Pumping Frequency $\omega^p/2\pi$ (\si{MHz}) & & $\omega_{\ket{B}}$ & \\
        \bottomrule
    \end{tabular}
\end{table}

\subsubsection{Calibration of Photon Number in Resonators}

\begin{figure}[htbp]
    \centering
    \includegraphics{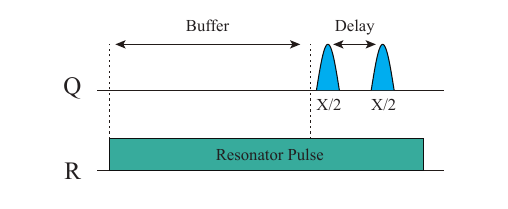}
    \caption{A pulse is applied to the resonator with a buffer time to ensure it reaches equilibrium before measurement. The photon number in the resonator can be inferred by measuring the shift in the qubit frequency as explained in the text.}
    \label{fig:meas_photon}
\end{figure}

Activating the Raman processes to engineer the desired dissipative channels requires applying coherent drives to the resonators. To calibrate the power of the drives, a Ramsey-like experiment is used to determine the stable average photon numbers $\Bar{n}_i$ in the resonators. The photon number $\Bar{n}_i$ strongly depends on the drive amplitudes ($\epsilon_i$) and frequencies ($\Delta_i^r$) via $\Bar{n}_i=\frac{|\epsilon_i^d|^2}{(\Delta_i^r)^2+(\kappa_i/2)^2}$, and it can be inferred by using the Ramsey-like experiment to measure the qubit frequency (\cref{fig:meas_photon}). This frequency includes an AC Stark shift of $2\Bar{n}_i\chi_i$ due to the qubit-resonator coupling \cite{gambetta_qubit-photon_2006}. The photon numbers corresponding to the resonator drives used in the experiment are listed in \cref{tab:res_drive_parameter}.

\begin{table}[htbp]
    \centering
    \tabcolsep=7mm
    \renewcommand{\arraystretch}{1.2}
    \caption{Photon numbers in the resonators and detunings of their drives in the two experiments.}
    \label{tab:res_drive_parameter}
    \begin{tabular}[c]{cccc}
        \toprule
        Bell state exp. & $Q_1$ & $Q_2$ & $Q_3$\\
        \midrule
        Photon number $\Bar{n}_i$ & 0.74 & 0.60 & \\
        Resonator drive detuning $\Delta^r_i$ & $\omega_{\ket{S}}-\omega_{\ket{T}}$ & $\omega_{\ket{S}}-\omega_{\ket{T}}$ & \\
        \midrule
        W state exp. & $Q_1$ & $Q_2$ & $Q_3$\\
        \midrule
        Photon number $\Bar{n}_i$ & 0 & 1.26 & 0.5\\
        Resonator drive detuning $\Delta^r_i$ & 0 & $\omega_{\ket{B}}-\omega_{\ket{W}}$ & $\omega_{\ket{A}}-\omega_{\ket{W}}$\\
        \bottomrule
    \end{tabular}
\end{table}

\subsection{Theoretical Analysis and Numerical Optimization}

\subsubsection{Error Analysis and Effective Model for Bell State Stabilization}

\begin{figure}[htbp]
    \centering
    \includegraphics{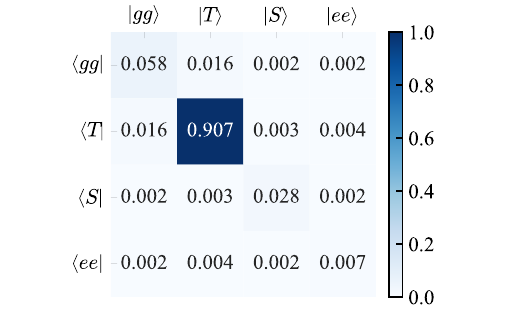}
    \caption{The density matrix of the target Bell state $\ket{T}$ in the eigenstate basis. The numbers and colors in the cells represent the element values.}
    \label{fig:supp_qst_dis}
\end{figure}

To analyze the errors that decrease the fidelity of the stabilized Bell state, we present its density matrix in the eigenstate basis, measured by quantum state tomography (QST) and averaged over the range from 4 to 10 $\mu$s with a 2 $\mu$s interval, as shown in \cref{fig:supp_qst_dis}. The fidelity of our target Bell state $\ket{T}$ is as high as $90.7\%$. The remaining population resides in the ground state $\ket{gg}$ ($5.8\%$), the $\ket{ee}$ state ($0.7\%$), and the orthogonal Bell state $\ket{S}$ ($2.8\%$).

\begin{figure}[htbp]
    \centering
    \includegraphics{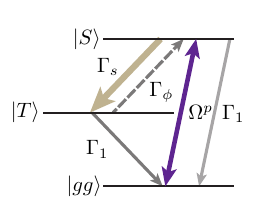}
    \caption{Effective three-level model for the experiment, aiding in understanding the fidelity of the stabilized state and methods to improve it.}
    \label{fig:supp_trilm}
\end{figure}

Although the eigenspace of the two-qubit system contains four states, the symmetry-selective pumping used in our work decouples the $\ket{ee}$ state from the rest, effectively reducing the system to a three-level model, as depicted in \cref{fig:supp_trilm}. The three-level model consists of the ground state $\ket{gg}$, the intermediate state $\ket{S}$, and the target state $\ket{T}$ to be stabilized. The system transitions from $\ket{S}$ to $\ket{T}$ at a rate of approximately $\Gamma_s$, mainly contributed by the dissipation engineered through the Raman process, along with the undesired reverse process from $\ket{T}$ to $\ket{S}$ at a low rate $\Gamma_\phi$. Additionally, both $\ket{T}$ and $\ket{S}$ can decay into the ground state $\ket{gg}$ at the same rate $\Gamma_1$ due to the finite $T_1$ of the qubits. Lastly, a resonant drive $\Omega^p$ is applied to pump the system from the ground state to $\ket{S}$. The dynamics of this driven-dissipative open system are described by a master equation:
\begin{equation}
    \begin{aligned}
        \dot{\rho} &= i[\rho, H]
        + \Gamma_1 \mathcal{D}(\ketbra{gg}{T})
        + \Gamma_1 \mathcal{D}(\ketbra{gg}{S})
        + \Gamma_s \mathcal{D}(\ketbra{T}{S})
        + \Gamma_\phi \mathcal{D}(\ketbra{S}{T}),
    \end{aligned}
\end{equation}
where $H = \frac{\Omega^p}{2}(\ketbra{gg}{S}+h.c.)$ represents the Hamiltonian of the model, while the remaining terms denote the dissipative channels. The steady-state solution is obtained by setting $\dot{\rho}=0$. Consequently, the steady-state fidelity $F$ of the target state $\ket{T}$ is expressed as:
\begin{equation}
F=\frac{(\Omega^p)^2\Gamma_s}{(\Omega^p)^2(2\Gamma_1+2\Gamma_\phi+\Gamma_s)+\Gamma_1^3+2\Gamma_1^2\Gamma_s+\Gamma_1\Gamma_s^2+\Gamma_1^2\Gamma_\phi+\Gamma_1\Gamma_s\Gamma_\phi}\label{Fidelity}.
\end{equation}
Given that in our experiment, the parameters satisfy $\Gamma_s\gg \Omega^p \gg \operatorname{max}\{\Gamma_\phi,\Gamma_1\}$, the terms containing $\Omega^p$ in the denominator dominate and \cref{Fidelity} can be further simplified to
\begin{equation}
F\approx\frac{\Gamma_s/2}{(\Gamma_1+\Gamma_\phi)+\Gamma_s/2}\label{eff_stable}.
\end{equation}
\cref{eff_stable} can be understood as follows. We take the state $\ket{T}$ as part A, while the rest of the subspace $\{\ket{gg},\ket{S}\}$ is part B. The effective decay rate from part A to part B is $\Gamma_{\rm{out}}=\Gamma_1+\Gamma_\phi$, while the rate from part B to part A is $\Gamma_{\rm{in}}=\Gamma_s/2$. The driven-dissipative system has a steady-state probability of $\Gamma_{\rm{in}}/(\Gamma_{\rm{in}} + \Gamma_{\rm{out}})$ of staying in part A. In the main manuscript, the experimental value $\Gamma_s^{\rm{exp}}$ is derived from the decay of $\ket{eg},\ket{ge}$, representing the total decoherence of the system, with $\Gamma_s^{\rm{exp}}=\Gamma_{\rm{in}} + \Gamma_{\rm{out}}$. Furthermore, $\Gamma_1$ can be easily obtained from separate single-qubit measurements, and $\Gamma_\phi$ is estimated as the average spin echo times of the two qubits. Consequently, the modified formula for fidelity, $(\Gamma^{\rm{exp}}_s - \Gamma_1 - \Gamma_\phi)/\Gamma^{\rm{exp}}_s$, yields a value of $91.7\%$, in good agreement with the experimental result of $90.7\%$ obtained via QST.

Finally, this simple model suggests two ways to improve fidelity: either increasing the dissipative rate from other states to the target state or decreasing the rate from the target state to other states (better qubit decoherence performance).

\subsubsection{Optimizing Parameters via Numerical Simulation}

To explore potential improvements in the stabilization process, we have performed numerical simulations for Bell state stabilization. For simplicity, the drive is applied only to the resonator of $Q_2$, activating the Raman process that induces the $\ket{S} \to \ket{T}$ transition in the simulation. In the following, we summarize the simulation results.

\begin{figure}[htbp]
    \centering
    \includegraphics{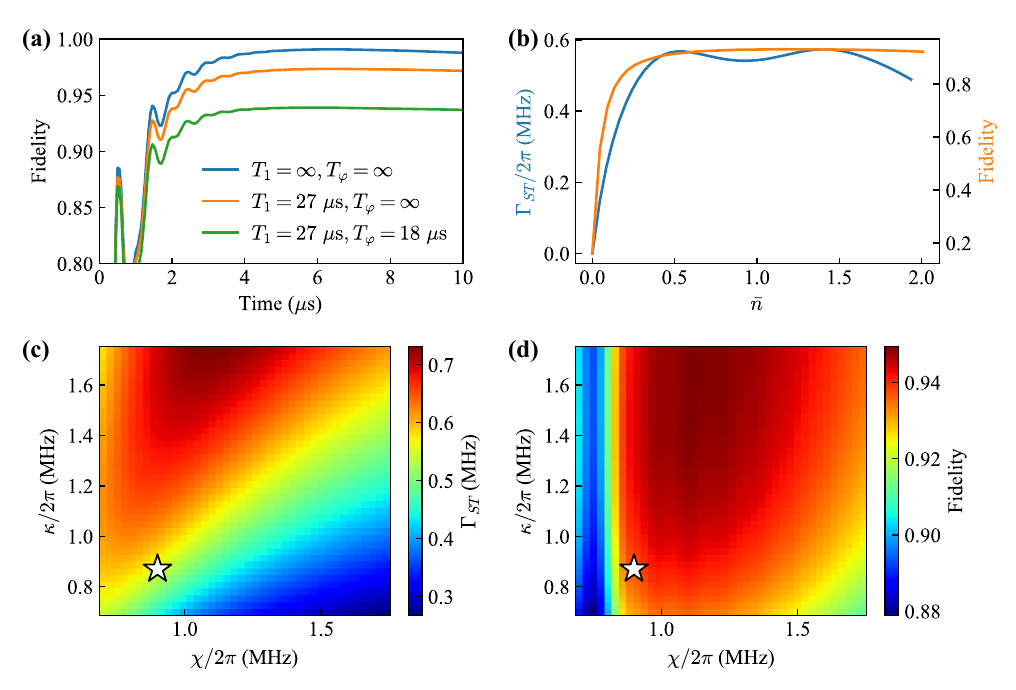}
    \caption{Simulating the impact of parameters on Bell state stabilization. (a) Dependence of the Bell state fidelity on decoherence parameters. (b) Dependence of the transition rate and fidelity on the photon number in the resonator. (c) and (d) The engineered dissipative rate $\Gamma_{ST}$ and stabilization fidelity as functions of $\chi$ and $\kappa$ with a fixed photon number of $0.74$ in the resonator. The star symbols correspond to the parameters used in our experiment.}
    \label{fig:supp_param_opt}
\end{figure}

1. As shown in \cref{fig:supp_param_opt}(a), the Bell state fidelity can reach up to $99\%$ if $T_1$ and $T_\phi$ of the qubits are infinite. However, a finite $T_1=\SI{27}\mu$s reduces the fidelity to $97\%$, and combining this with a $T_\phi=\SI{18}\mu$s results in $94\%$. Therefore, qubits with better $T_1$ and $T_\phi$ are always preferred, along with a faster engineered dissipative channel to counteract the $T_\phi$ process.

2. \cref{fig:supp_param_opt}(b) shows that both the engineered dissipative rate $\Gamma_{\ket{S}\to\ket{T}}=\Gamma_{ST}$ and the fidelity of the stabilized Bell state are strongly dependent on the average photon number $\bar{n}$ in the resonator. Optimization of the excited photon number inside the resonator is needed for higher fidelity and faster stabilization.

3. Although the two parameters of dispersive shift $\chi$ and dissipative rate $\kappa$ of the resonators are fixed at $\kappa \approx \chi$ in our sample, we extract both the engineered dissipative rate $\Gamma_{ST}$ and the fidelity of the stabilized Bell state depending on $\chi$ and $\kappa$ from the master equation simulation with a fixed average photon number of $0.74$ in the resonator, as shown in \cref{fig:supp_param_opt}(c) and (d). Numerical simulations indicate that a faster $\kappa$ and suitable $\chi$ can further enhance stabilization performance, including speed and fidelity. Furthermore, a faster $\kappa$ is also beneficial for fast readout in quantum error correction\cite{Krinner2022}.

\subsection{Bell state Stabilization from Different States}

\begin{figure}[htbp]
    \centering
    \includegraphics{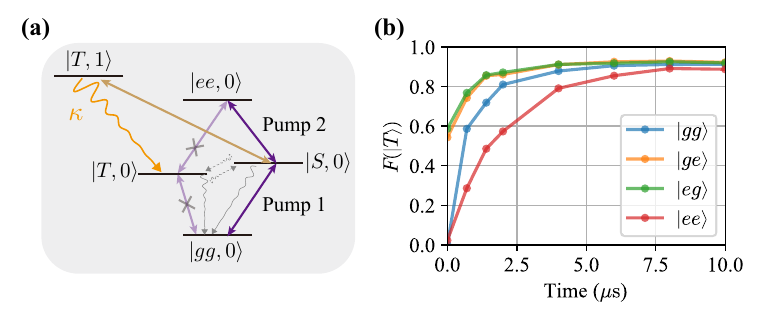}
    \caption{(a) Energy level diagram of the system with a Raman process using two drives. Pump 1 excites the transition $\ket{gg}\leftrightarrow\ket{S}$ as discussed in the main manuscript. An additional Pump 2 activates the transition $\ket{ee}\leftrightarrow\ket{S}$ while inhibiting the transition $\ket{T}\leftrightarrow\ket{gg}$. With these drives, all states in the $1$-excitation manifold can be stabilized to the state $\ket{T}$. (b) The fidelity of the Bell state $\ket{T}$ for four product initial states as a function of time.}
    \label{fig:full_stabilization}
\end{figure}

In the main manuscript, we demonstrated that the two-qubit system can be stabilized into the target Bell state when initialized in $\ket{gg}$, $\ket{ge}$, or $\ket{eg}$, while $\ket{ee}$ is decoupled. Here, we show that stabilizing the system from an initial state of $\ket{ee}$ is also possible. In addition to the resonant drive connecting the ground state $\ket{gg}$ and $\ket{S}$ (labeled as Pump 1 in \cref{fig:full_stabilization}(a)), as discussed in the main manuscript, we further apply another resonant drive to couple $\ket{ee}$ with $\ket{S}$, labeled Pump 2 in \cref{fig:full_stabilization}(a). The experimental results are depicted in \cref{fig:full_stabilization}(b). While the system can be stabilized into the target Bell state $\ket{T}$ from the initial state $\ket{ee}$, the stabilization speed is slower compared to other cases. The reason for this difference needs further investigation. We also note that another feasible approach to stabilizing the system from a general initial state is to first unconditionally reset all qubits to their ground states\cite{Magnard_reset,Zhou2021}, followed by the same strategy used in the main manuscript.

\newpage

\nocite{*}
\bibliography{SupplementalMaterial}

%% file: preamble.tex
\usepackage{graphicx}
\usepackage{amsmath}
\usepackage{color}
\usepackage{float}
\usepackage{epstopdf}
\usepackage{soul,xcolor}
\usepackage[utf8]{inputenc}
\usepackage[colorlinks=true,citecolor=blue,linkcolor=magenta]{hyperref}
\usepackage{verbatim}
\usepackage{lipsum}
\usepackage{physics}
\usepackage{siunitx}
\usepackage{dcolumn}
\usepackage{bm}
\usepackage{booktabs}
\usepackage{svg}

\usepackage{hyperref}
\usepackage{cleveref}
\usepackage{verbatim}
\crefname{figure}{Fig.}{Figs.}
\crefname{equation}{Eq.}{Eqs.}
\crefname{table}{Table}{Tables}

%% file: author.tex
\author{Changling Chen}
\altaffiliation{These authors contributed equally to this work.}
\affiliation{Shenzhen Institute for Quantum Science and Engineering, Southern University of Science and Technology, Shenzhen 518055, China}
\affiliation{Guangdong Provincial Key Laboratory of Quantum Science and Engineering, Southern University of Science and Technology, Shenzhen, 518055, China}
\author{Kai Tang}
\altaffiliation{These authors contributed equally to this work.}

\affiliation{Guangdong Provincial Key Laboratory of Quantum Science and Engineering, Southern University of Science and Technology, Shenzhen, 518055, China}
\affiliation{Department of Physics, Southern University of Science and Technology, Shenzhen 518055, China}

\author{Yuxuan Zhou}

\affiliation{Guangdong Provincial Key Laboratory of Quantum Science and Engineering, Southern University of Science and Technology, Shenzhen, 518055, China}
\affiliation{Department of Physics, Southern University of Science and Technology, Shenzhen 518055, China}

\author{KangYuan Yi}

\affiliation{Guangdong Provincial Key Laboratory of Quantum Science and Engineering, Southern University of Science and Technology, Shenzhen, 518055, China}
\affiliation{Department of Physics, Southern University of Science and Technology, Shenzhen 518055, China}

\author{Xuan Zhang}

\affiliation{Guangdong Provincial Key Laboratory of Quantum Science and Engineering, Southern University of Science and Technology, Shenzhen, 518055, China}
\affiliation{Department of Physics, Southern University of Science and Technology, Shenzhen 518055, China}
\author{Xu Zhang}
\affiliation{Guangdong Provincial Key Laboratory of Quantum Science and Engineering, Southern University of Science and Technology, Shenzhen, 518055, China}
\affiliation{Department of Physics, Southern University of Science and Technology, Shenzhen 518055, China}

\author{Haosheng Guo}
\affiliation{Guangdong Provincial Key Laboratory of Quantum Science and Engineering, Southern University of Science and Technology, Shenzhen, 518055, China}
\affiliation{Department of Physics, Southern University of Science and Technology, Shenzhen 518055, China}

\author{Song Liu}
\affiliation{Shenzhen Institute for Quantum Science and Engineering, Southern University of Science and Technology, Shenzhen 518055, China}
\affiliation{Guangdong Provincial Key Laboratory of Quantum Science and Engineering, Southern University of Science and Technology, Shenzhen, 518055, China}
\affiliation{International Quantum Academy, Shenzhen 518048, China}

\affiliation{Shenzhen Branch, Hefei National Laboratory, Shenzhen 518048, China}

\author{Yuanzhen Chen}
\email{chenyz@sustech.edu.cn}
\affiliation{Shenzhen Institute for Quantum Science and Engineering, Southern University of Science and Technology, Shenzhen 518055, China}

\affiliation{Guangdong Provincial Key Laboratory of Quantum Science and Engineering, Southern University of Science and Technology, Shenzhen, 518055, China}
\affiliation{Department of Physics, Southern University of Science and Technology, Shenzhen 518055, China}

\author{Tongxing Yan}
\email{yantx@sustech.edu.cn}
\affiliation{Shenzhen Institute for Quantum Science and Engineering, Southern University of Science and Technology, Shenzhen 518055, China}
\affiliation{Guangdong Provincial Key Laboratory of Quantum Science and Engineering, Southern University of Science and Technology, Shenzhen, 518055, China}
\affiliation{International Quantum Academy, Shenzhen 518048, China}

\author{Dapeng Yu}
\affiliation{Shenzhen Institute for Quantum Science and Engineering, Southern University of Science and Technology, Shenzhen 518055, China}

\affiliation{Guangdong Provincial Key Laboratory of Quantum Science and Engineering, Southern University of Science and Technology, Shenzhen, 518055, China}
\affiliation{Department of Physics, Southern University of Science and Technology, Shenzhen 518055, China}
\affiliation{International Quantum Academy, Shenzhen 518048, China}
\affiliation{Shenzhen Branch, Hefei National Laboratory, Shenzhen 518048, China}